\documentclass[pmlr,twocolumn,10pt]{jmlr} 

\usepackage{booktabs}
\usepackage{siunitx}

\usepackage[switch]{lineno}

\theorembodyfont{\upshape}
\theoremheaderfont{\scshape}
\theorempostheader{:}
\theoremsep{\newline}

\jmlrvolume{Page}
\jmlryear{2024}
\jmlrsubmitted{}
\jmlrpublished{}
\jmlrworkshop{Final revision: 2024, April 15th @ ArXiv} 

\title[Luffina C. H. 2024, Greedy-optimized Approach for IBI/HRV Daily Monitoring using Wearable PPG]{Greedy-optimized Approach for Interbeat Interval and Heart Rate Variability Daily Monitoring using Wearable PPG}

 \author{%
  \Name{Luffina C. Huang} \Email{luffina.c.huang@rice.edu}\\
  \addr Rice University, Electrical and Computer Engineering
}

\makeatletter
\newcommand{\algorithmfootnote}[2][\footnotesize]{%
  \let\old@algocf@finish\@algocf@finish
  \def\@algocf@finish{\old@algocf@finish
    \leavevmode\rlap{\begin{minipage}{\linewidth}
    #1#2
    \end{minipage}}%
  }%
}
\makeatother

\begin{document}

\maketitle

\begin{abstract}
Continuous monitoring of heart rate variability (HRV) provides insights in cardiovascular health. Wearable Photoplethysmography (PPG) assures convenient measurement of HRV. PPG, however, is susceptible to motion artifacts, considerably deteriorating the accuracy in estimation. In this study, a greedy-optimized approach is proposed for attaining high accuracy of interbeat intervals (IBIs) estimation from PPG signals collected during intensive daily activities. Utilizing the fact of continuity in heartbeats, the IBI estimation is converted into the shortest path problem in a directed acyclic graph, where candidate heartbeats from motion-contaminated PPG are regarded as vertices. The approach exploits a convex penalty function to optimize weight assignment in the shortest path calculation and a greedy fusion method to strengthen the selection process of optimal IBIs. Results achieve correlation of 0.96 for IBI estimation with the improvement of 58.4\% in percentage error on the single-channel PPG signal from 2015 IEEE Signal Processing Cup. It also achieves correlation of 0.98 with percentage error of 2.2\% in the two-channel PPG signal. Estimated and true HRV parameters are highly correlated. The approach is further validated on the PPG-DaLiA dataset with high correlation and low percentage error for IBI and HRV estimation in two daily activities, indicating the robustness of the proposed technique.
\end{abstract}

\paragraph*{Data Availability}

This paper uses two publicly available datasets, the 2015 IEEE Signal Processing Cup \citep{10} and the PPG-DaLiA \citep{29}.

\section{Introduction}
\label{sec:intro}

Continuous cardiovascular activity monitoring is increasing in popularity through emerging wearable devices, benefiting the quality of healthcare and self-health management. Average HR, the number of heart beats per minute, is one of the vital signs routinely monitored by healthcare providers and serves as an important indicator of cardiovascular health, such as hemodynamic stability and heart rhythm. Heart rate variability (HRV), however, can provide integrated information of the cardiovascular system and the autonomic nervous system, thus becoming a crucial physiological parameter in continuous cardiovascular monitoring. Interbeat intervals (IBIs) are the time elapsed between two successive heart beats. HRV quantifies the variability of IBIs in a certain period of time and is widely used as a crucial indicator in health research and clinical practice. HRV parameters can be used to evaluate the sympathetic and parasympathetic activity of the autonomic nervous system, such as respiration, physical stress and mental load ~\citep{1}. Further, reduced HRV, i.e., reduced level of beat-to-beat heart rate fluctuations, is independently associated with a 32–45\% increased risk of first fatal and non-fatal cardiovascular disease (CVD) events and a prognostic factor with higher mortality in patients with CVD events~\citep{1,2,3}. 

IBIs and HRV can be derived from both Electrocardiography (ECG) and Photoplethysmography (PPG)~\citep{1}. Traditionally, the gold standard of IBIs and HRV measurements is multi-lead ambulatory ECG. Although ambulatory ECG provides out-of-hospital monitoring, it requires specialized setups and multiple electrodes to the chest. PPG-based wearables are convenient for continuous IBIs and HRV monitoring in daily life, serving as an alternative to the standard ECG~\citep{1,3,4}. PPG is an optical biomonitoring technique to measure blood volumetric changes at fingers and wrists~\citep{5}. Featured physiological parameters related to the cardiopulmonary system could be estimated by PPG, such as blood oxygen saturation, average heart rate (HR) and respiratory rate~\citep{6}. Studies have shown the IBIs and HRV parameters derived from PPG wearables are highly associated with those derived from ECG signals in stationary conditions with correlation in the range from 0.85 to 0.99 ~\citep{4}. Motion artifacts, nevertheless, are an inherent problem if employing wearable PPG in daily or intensive activities, degrading the accuracy of IBI/HRV estimation as the level of physical activity increases. Denoising PPG signals through general filtering techniques is difficult because the frequency spectrum of motion artifacts (0.01 - 10 Hz) overlaps with the normal frequency of PPG signal (0.5 - 5 Hz)~\citep{8}. IBI/HRV exhibits inconsistency over time with overestimated fluctuating patterns and is challenging to obtain from PPG wearables during intensive activities ~\citep{19}. Furthermore, prevailing commercial wearable devices favor single-channel PPG sensors~\citep{3}. Therefore, techniques for accurate IBI and HRV estimation from single-channel motion-contaminated PPG are needed.

In this study, a greedy-optimized approach is proposed to tackle this unmet need, which transforms IBI estimation into a shortest path problem in directed acyclic graph subsequently combined with a greedy fusion method for morphological features extracted from motion-contaminated PPG. By using the physiological property of the temporal continuity of heartbeats (i.e., the end of one heartbeat is the start of the next heartbeat), a directed acyclic graph is constructed, where the vertices represent the feature candidates (i.e., heartbeats) and the edges represent the candidate IBIs. Shortest path algorithm is then used to select true heartbeats from candidates. The proposed convex penalty function for weight assignment is designed to augment the power of the shortest path algorithm with increments of the accuracy in IBI estimation. Subsequently, the greedy fusion method is designed for optimizing the selection process of estimated IBIs from three morphological features, systolic peaks, maximum slope, and onset. This greedy fusion technique soothes the inherent overly fluctuating patterns of estimated IBIs from noisy PPG signals, which enhances the performance of IBI/HRV estimation from single-channel PPG collected during intensive activities. It allows the adaptiveness with single-contact commercial wearable devices, which have, in general, limited computational capacity. Note that this approach could be further adopted with multi-channel PPGs through observed properties of morphological features~\citep{9}. The contributions of this article are as the follow: 
\begin{enumerate}
\item A greedy-optimized approach is proposed to attain high accuracy of IBI/HRV estimation from motion-contaminated PPG. Convex penalty function optimizes weight assignment to augment the power of the shortest path algorithm. Greedy fusion method strengthens the selection process of optimal IBIs. 
\item IBI and HRV estimation performance are validated on two public datasets, IEEE Signal Processing Cup and PPG-DaLiA, which have noisy PPG from wearables collected during intensive daily activities, indicating the robustness of proposed techniques.
\end{enumerate}

\section{Related Works}

\begin{figure*}[ht]
\vskip 0.0in
\begin{center}
\centerline{\includegraphics[width=2.1 \columnwidth]{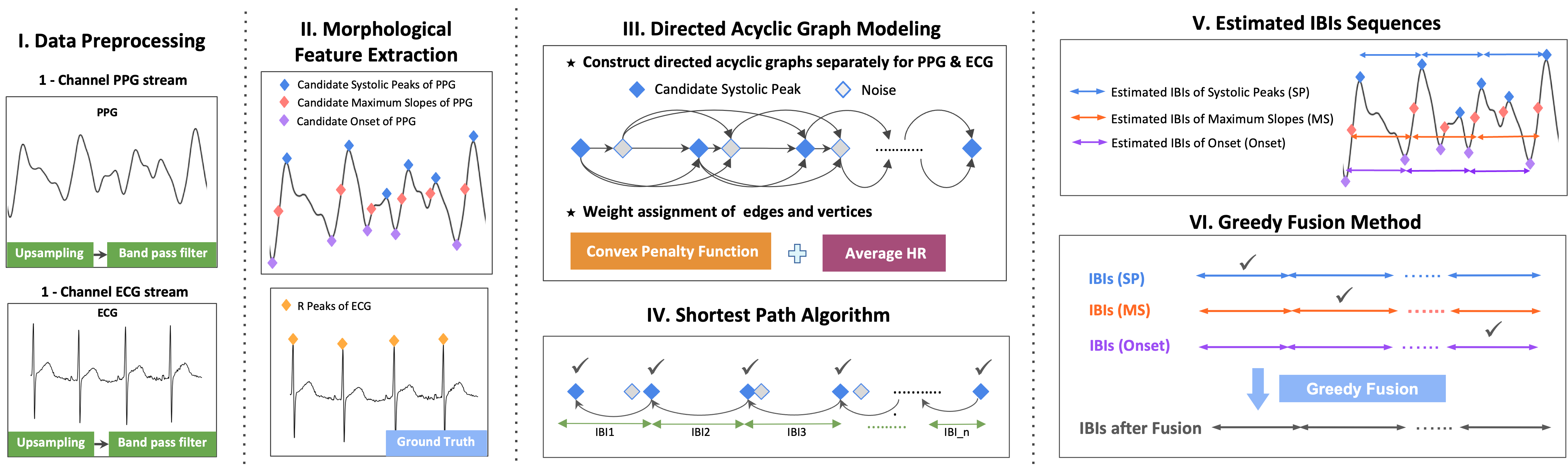}}
\end{center}
\vskip -0.4in 
\caption{Overview of greedy-optimized approach in attaining estimated IBIs from noisy PPG.}
\label{Overview}
\end{figure*}

\paragraph{Heart Rate Variability Estimation.} IBI/HRV have extensive physiological applications in clinical practice, but it is challenging to attain accurate IBIs from wearable PPG sensors. Initially, some studies demonstrate the feasibility of IBI and HRV estimation using wrist-worn PPG sensors on post-anesthesia patients and healthy volunteers during sleep~\citep{15,16}. Although these studies shows small absolute errors of IBIs and HRV parameters between wrist-worn PPG and standard ECG, most PPG signals are not distorted by motion artifacts. Although one study shows the satisfying association with correlation ranging from 0.74 to 0.88 between wrist-worn PPG sensors and standard ECG in HRV parameters in rest condition, the correlation is degraded to 0.42 - 0.67 when subjects were talking~\citep{17}. Another work has benchmarked the HRV parameters for different ages and genders using a dataset of 8 million users. It applies noise spikes cleaning algorithms and achieves high correlation of 0.97 in HRV parameters during a randomly selected 24 hours period~\citep{3}. However, above two studies eliminate all PPG signals that are corrupted with motion artifacts from their HRV analysis. A study explores accurate R peaks detection in noisy ECG signals, which applies the IncResU-Net, a fully convolutional Encoder-Decoder architecture, to detect R-peak from the ECG. The model provides good performance with precision of 0.82 in R-peak detection in ECG with noise level up to 0 dB. That work, however, does not investigate into motion-contaminated PPG signals~\citep{18}. A study leverages graph theory and signal-switching fusion method on single-channel motion-corrupted PPGs, which presents a satisfying correlation of 0.82 to 0.89 between the PPG sensors and standard ECG, but accuracy evaluation is not reported for both IBI and HRV estimation ~\citep{19}. Another study proposes a graph-based model for multi-channel PPG signals, which presents enhanced performance than those using the single-channel PPG signals in IBI and HRV estimation during intensive activity. Nevertheless, the overestimated fluctuating patterns of estimated IBIs have still remained and the model could not adapt with single-contact wearables ~\citep{9}.

\paragraph{Fusion Methods of Physiological Signals.} Fusion approaches have been explored to enhance the accuracy of heartbeats detection by incorporating the information across different physiological signal modalities or multiple morphological features. One fusion approach is signal switching, where candidate fiducial points from a signal modality with the best signal quality are selected as final fiducial points in a certain segment. \citet{20} uses the sample entropy to assess the noise content in multiple signal modalities, such as ECG and arterial blood pressure (ABP) signals, and switch between them to enhance the accuracy of heartbeat detection. \citet{19} obtains the best set of IBI arrays from three PPG morphological features by selecting those segments with minimal standard deviation of IBI subarray. Another fusion approach related to voting method, where candidate fiducial points detected in each signal modality cast a vote to select final fiducial points for a certain segment. In the majority voting, the fiducial points that have most agreement among different signal modalities are selected as the final fiducial points~\citep{21}. Furthermore, the vote could be weighted by the signal quality index or other evaluation metrics to select fiducial points with best quality~\citep{22}. Other fusions are based on probabilistic models. In a study, authors employ Bayesian Network to model the relationship between the ECG, ABP and classification for hidden states in a Hidden Markov Model ~\citep{23}.  

\section{Methods}
The overview of attaining accuracy IBI estimation from motion-contaminated PPG using greedy-optimized approach is shown in Fig.~\ref{Overview}, which consists of data preprocessing, feature extraction, directed acyclic graph modeling, shortest path algorithm and greedy fusion method. In the preprocessing stage, PPG signals are upsampled and filtered. Then, PPG morphological features of a cardiac cycle are extracted as candidate fiducial points to represent the potential heartbeats. Firstly, a directed acyclic graph is constructed where the vertices represent all potential heartbeats, and the edges represent all the candidate IBIs. Secondly, true heartbeats are differentiated from the noise spikes using a shortest path algorithm with convex penalty function. The time differences of two consecutive heartbeats selected by the shortest path algorithm are regarded as estimated IBIs. Finally, a greedy fusion method that utilizes the complementary information from three morphological features (systolic peaks, maximum slope and onset) further improves the accuracy of IBI estimation. Note that both shortest path calculation and greedy fusion method are generalized and could be adopted into different physiological signal modalities, such as ECG and arterial blood pressure waveform.

\begin{figure}[ht]
\vskip 0.0in
\begin{center}
\centerline{\includegraphics[width=0.8 \columnwidth]{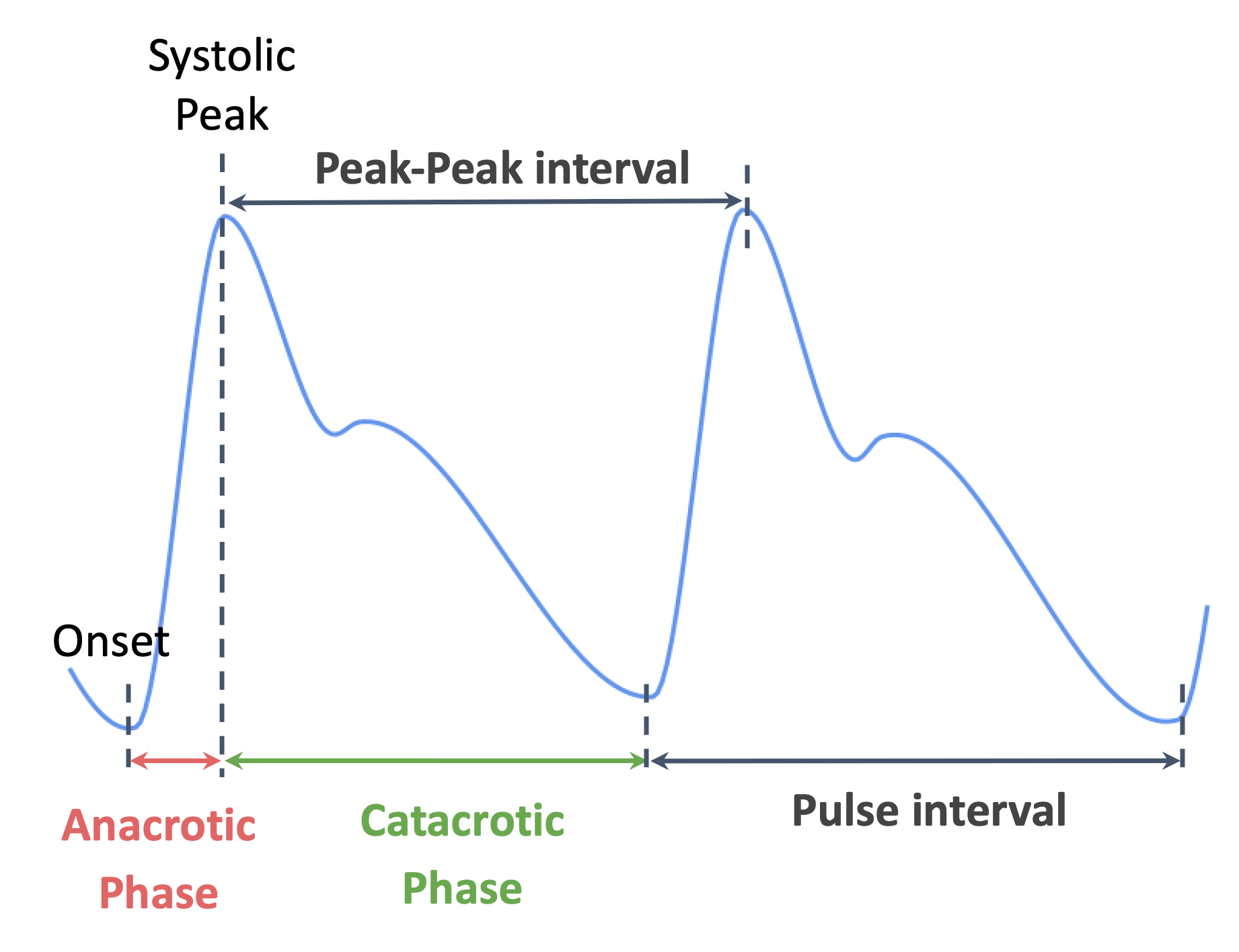}}
\end{center}
\vskip -0.3in
\caption{Standard PPG waveforms}
\label{PPGwaveforms}
\end{figure}

\subsection{Morphological Feature Extraction}
In ECG signals, the R-peak is the most prominent peak of R-wave in the QRS complex. The time elapsed between two consecutive R-peaks, called R-R interval, is the most common and standard way to calculate IBIs.  To strengthen the R-peak detection, ECG signals are preprocessed by a continuous wavelet transform (CWT) using Mexican hat with a center frequency of 0.25Hz~\citep{19}. The peaks detected from the wavelet are regarded as ground truth R-peaks and the calculated IBIs from those peaks are regarded as the ground truth. 

The PPG waveform of a cardiac cycle is commonly divided into two phases, anacrotic (systolic) phase  and catacrotic phase as shown in Fig.~\ref{PPGwaveforms}. Systolic peaks, maximum slopes and onset points are important PPG morphological features that characterize the systolic waveform of a cardiac cycle~\citep{24}. The time elapsed between two consecutive systolic peaks in PPG signals is referred to as Peak-Peak interval, whereas the time elapsed between the onset and the end of the PPG waveform is referred to as Pulse interval, as shown in Fig.~\ref{PPGwaveforms}. Some studies observe that the Peak-Peak interval in PPG signals is highly correlated with the R-R interval in ECG signals~\citep{25}. Other studies show that HRV from the Pulse interval in PPG signals are highly correlated with HRV from R-R intervals in ECG signals. Both Peak-Peak interval and Pulse interval have been used to detect heart rate and HRV in stationary condition ~\citep{26,27}. Maximum slope, which is centered between systolic peak and onset in PPG signals, has also been applied to HRV estimation in intensive physical activity ~\citep{19}. Therefore, these three PPG morphological features are used for IBI estimation in this study. Following steps are the process of extracting these features, which are regarded as candidate fiducial points.  First, the filtered PPG signals are smoothed by a 5th order smoothing spline. Then, a general peak detection algorithm in SciPy~\citep{28} detects the local maxima of PPG signals to obtain systolic peak candidates of PPG. Then, I use the same strategy to extract the local maximum of the first derivative and second derivative of PPG signals as maximum slope candidates and onset candidates, respectively.

\subsection{Directed Acyclic Graph Moedling with Convex Penalty Function} 
In this step, morphological features start converting into a directed acyclic graph. Initially, three directed acyclic graphs are constructed using candidate fiducial points extracted from three morphological features (systolic peaks, maximum slopes and onset), respectively. In the graph, vertices are marked as the candidate fiducial points while edges are designed as candidate IBIs. Vertices are denoted as $v_i,i=1,\dots,n$, where $n$ is the total number of vertices in the graph and their values are equal to their timestamp. A time interval $t_i$ prior to each vertex $v_i$ with the range of 1.5 folds of its average IBI is considered to identify neighbors of $v_i$, where $(IBI_{avg})_i=6000/ (HR_{avg})_i $ (ms). Edges are formed between vertex $v_i$ and its neighbors and denoted as $e_{ij}$, where vertices $v_j$ are $v_i$’s neighbors. Average heart rates, $HR_{avg}$, are estimated from PPG1 using the WFPV algorithm, where $(HR_{avg})_i$ of vertex $v_i$ is equal to the average heart rate of the 8-second PPG window which is closest to $v_i$ ~\citep{11}. Following the above steps, the graph construction of the single-channel PPG signal is completed.

The shortest path algorithm is used as the graph search algorithm to attain the correct IBI path. The weight of each edge is assigned based on their deviation from the average IBI. The weight of each vertex is accumulated from the start vertex to the current one. Therefore, an effective penalty function is crucial for assigning edge weights of the graph. A convex penalty function proposed in~\citep{9} showed the better results for IBIs estimation and is used in this study:
\begin{align}
    v'_i &=v_i-(IBI_{avg})_i, \hspace{0.5cm} i = 2, \dots ,n  \\
    d_{ij} &=|v_j-v'_i|, \hspace{0.5cm} j = i-1, \dots ,i-m_i  \\
    w_{ij} &= \lambda {d_{ij}}^x, \hspace{0.5cm} x \in \mathbb{N} 
\end{align} %
where $v'_i$ is marked as the expected previous vertex of $v_i$. $v_j$  are $v_i$’s neighbors and $m_i$ is the total number of neighbors of vertex $v_i$. $d_{ij}$ is assigned as the time difference of the two connected vertices. The weight $w_{ij}$ of the edge that connects $v_i$ to neighbor $v_j$ is calculated by raising the $d_{ij}$ to the power of $x$ with a constant parameter $\lambda$, where $x$ is assigned as 2 in this study to attain strong convexity. Note that this directed acyclic graph modeling can also be adapted to multi-channel PPGs, by leveraging the phenomenon that false fiducial points included by noise would have a bigger time gap between PPG1 and PPG2 than the true fiducial points, vertices from two channels are concatenated and sorted by timestamps to model the multi-channel PPGs into a directed acyclic graph~\citep{9}.

\subsection{Shortest Path Algorithm} 
Since the end of each heartbeat is the beginning of its next heartbeat and they are continuous in time domain, the shortest path is used to select the fiducial points that correspond to true heartbeats. After constructing the weighted graph, the shortest path algorithm is applied on this graph and then the path with the least total weight is chosen. The weight of vertex $v_i$ is assigned by finding the minimum of the weight of previous neighbors plus the weight of edges connected between them. The previous neighbor that contributes to the minimum is selected as the previous vertex of $v_i$ and denoted as $pn_i$, which is shown in Algorithm~\ref{alg:shortestPath}.

\begin{algorithm2e}[h]
 \caption{Shortest Path Calculation by Convex Penalty Function}
 \label{alg:shortestPath}
\KwIn{Candidate fiducial points from 1 channel PPG signal}  \par  
\KwOut{the set of chosen vertices $\mathcal{V}_{chosen}$}
Construct a directed acyclic graph by timestamp of candidate fiducial points to form the vertex set $\mathcal{V} = [v_1, v_2,...,v_n]$, where $n$ is total number of vertices \\
\tcp{Neighbor selection for vertex $v_i$}
 \For{$v_i \in \mathcal{V}$}{ 
   $j = i-1$ \\
   \While{($v_i - v_j) < 1.5*(IBI_{avg})_i$}{  
     $e_{ji} = |v_i - v_j|$ \\
     $\mathcal{N}_i \gets v_j$ \\
     $j = j - 1$
    } 
     $m_i = i-j +1$ 
}
\tcp {Weight assignment for vertices} 
  \For{i = 2 \KwTo n} {
    $v'_i = v_i-(IBI_{avg})_i$\\
   \For{j = $i-1$ \KwTo $i-m_i$}{
        $d_{ij} =|v_j-v'_i|$ \\
        $w_{ij} = \lambda {d_{ij}}^2$   } 
    $w_i$ $=$ ${\mathrm{min}}(w_{ij} + w_j)$, $for$ $v_j \in \mathcal{N}_i$\\ 
    $pn_{i}$ $=$ $\underset{v_j \in \mathcal{N}_i}{\mathrm{argmin}} (w_{ij} + w_j)$, $for$ $v_j \in \mathcal{N}_i$ 
   } 
    $v_{dst}$ $=$ $\underset{v_k \in [\mathcal{N}_n, v_n]}{\mathrm{argmin}} w_k$\\ 
    Backward search from $v_{dst}$ to select the vertices to form $\mathcal{V}_{chosen}$ \\
    \Return $\mathcal{V}_{chosen}$
 \algorithmfootnote{$\mathcal{N}_i$ denotes the set of neighbors of vertex $v_i$. $m_i$ denotes the number of vertices in $\mathcal{N}_i$. $w_i$ denotes the weight of vertex $v_i$. $pn_i$ denotes the chosen previous vertex which contributes to the minimal weight for $v_i$}

 \end{algorithm2e}

\subsection{Greedy Fusion Method} 
One IBI sequence is produced from one of three morphological features, respectively. Consecutive IBIs, however, would not be estimated precisely from motion-corrupted PPG. Those beat-to-beat IBI plots, as shown in Fig.~\ref{IBIplot}, depict that estimated IBIs have overestimated fluctuating patterns, as compared to the true IBIs from ECG, due to the difficulty of extracting the true fiducial points from highly-distorted PPG signals. In this study, a greedy fusion method is proposed to tackle this challenge.

Due to the fact that the onset feature represents the beginning of a cardiac activity, IBIs sequence from the onset feature is selected as the baseline for segmentation. Firstly, the IBIs sequence of the onset feature is divided into q segments where each segment contains three consecutive IBIs. The timestamps of this segmentation are used as references to guide the segmentation of IBIs sequences for maximum slope and systolic peak features. For each segment, if the starting points of IBIs from maximum slope and systolic features are within this segment, these IBIs are included as the candidate IBIs. Secondly, based on physiological phenomenons that true IBIs are expected to be close to their average IBIs and would not have drastic changes for short time, hence, an objective function as Equation~(\ref{objective}) is proposed for the greedy fusion method to find local optimum IBIs in each segment.
\begin{equation}
\hat{j},\hat{k},\hat{l} = \underset{ j \neq k \neq l}{\mathrm{argmin}}\{ |i_{Cp}^j - i_{Ap}^1 | + |i_{Cp}^k-i_{Ap}^2 | + |i_{Cp}^l - i_{Ap}^3|\} 
\label{objective}
\end{equation}
The process of greedy fusion method on three morphological features is shown in Algorithm~\ref{alg:fusion}. First, candidates from morphological features need to be identified. The IBIs sequence generated from onset are denoted as $I_O$, IBIs sequence from systolic peak as $I_S$, and IBIs sequence from maximum slope as $I_M$. The $p$th segment of $I_O$, $I_S$, and $I_M$ are denoted as $I_{Op}$, $I_{Sp}$, and $I_{Mp}$. The individual IBIs in $I_{Op}$ are denoted as $i_{Op}^1,\dots ,i_{Op}^k$, where $k$ must be equal to 3 for segments in $I_O$, but $k$ can be any integer number close to 3 for segments in $I_S$ and $I_M$.

\begin{figure}[h]
\begin{center}
\centerline{\includegraphics[width=1 \columnwidth]{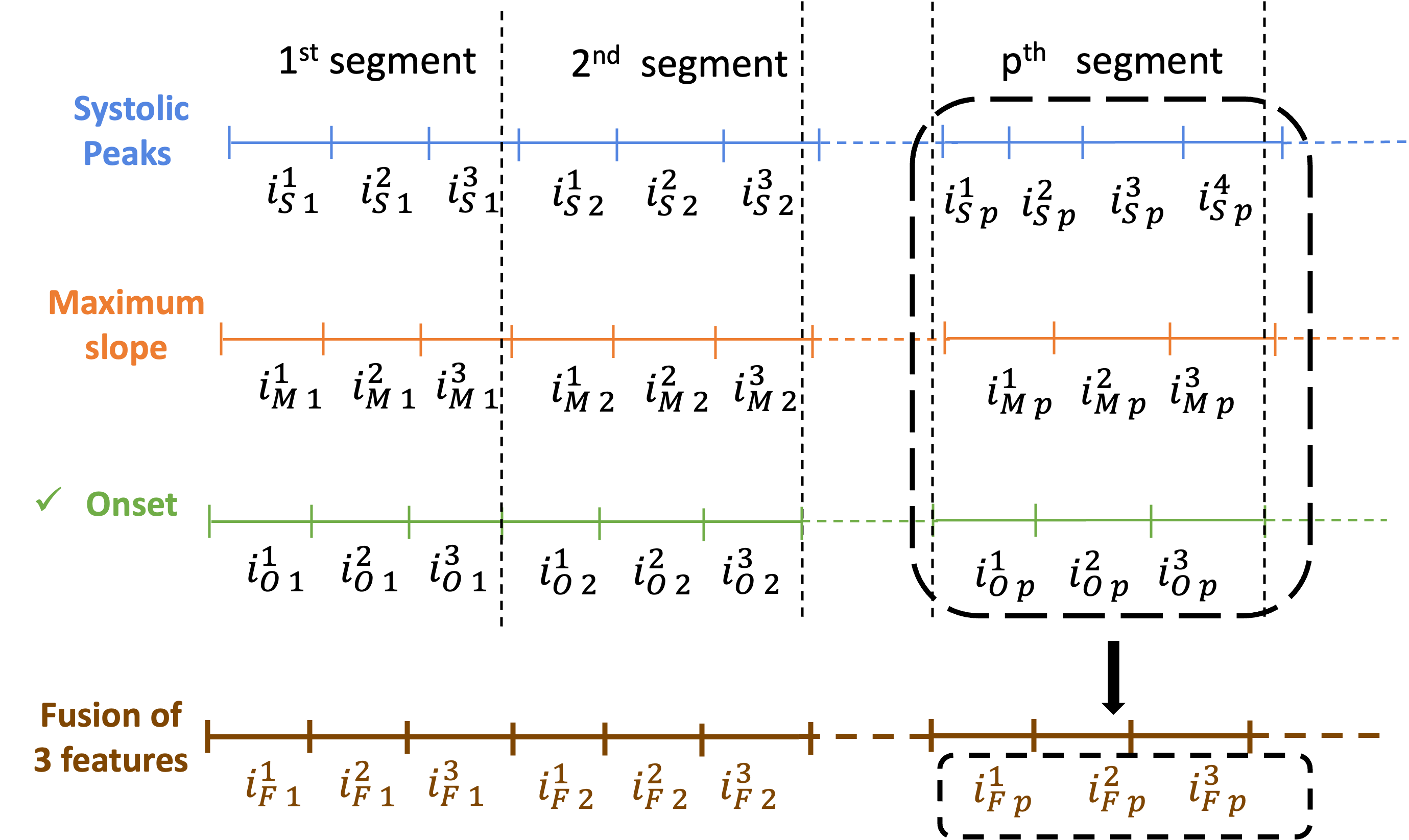}}
\end{center}
\vskip -0.3in
\caption{Greedy fusion method for various shortest paths utilizing morphological features.}
\label{GreedyFusion}
\end{figure}

Then, greedy fusion method would go through every segment to find local optimum IBIs. Take Fig.~\ref{GreedyFusion} as example, the set of candidate IBIs in the $p$th segment, denoted as $I_{Cp}$, is the union of $I_{Op}$, $I_{Sp}$, and $I_{Mp}$. The sequence of average IBIs in the pth segment are denoted as $I_{Ap}$. The starting points of four IBIs from $I_S  (i_{Sp}^1,i_{Sp}^2,i_{Sp}^3,i_{Sp}^4)$ and three IBIs from $I_M  (i_{Mp}^1,i_{Mp}^2,i_{Mp}^3 )$ are within the pth segment. Candidate IBIs in the $p$th segment, $I_{Cp}$, is composed of $\{i_{Sp}^1,i_{Sp}^2,i_{Sp}^3,i_{Sp}^4,i_{Mp}^1,i_{Mp}^2,i_{Mp}^3,i_{Op}^1,i_{Op}^2,i_{Op}^3 \}$. The three IBIs in $I_{Cp}$ that minimize the absolute error are chosen as shown in Equation~(\ref{objective}) and are concatenated into the final IBIs sequence, $I_F$. After iterating $q$ segments to obtain the complete $I_F$, feasible optimized solutions are regarded as final estimated IBIs from motion-contaminated PPG.

{
\begin{algorithm2e}[h]
 \caption{Greedy Fusion Method}
 \label{alg:fusion}
\KwIn{The IBIs sequences from systolic peaks, maximum slope, and onset features, ie., $I_S$, $I_M$, and $I_O$. The average IBIs sequences, $I_A$}  \par  
\KwOut{The final IBIs sequence $I_F$}

Divide $I_O$ into $q$ segments such that each segment has 3 IBIs. \\
Divide $I_S$ and $I_M$ based on $I_O$ segmentation. \\
$I_{F} = \varnothing$ \\
\For{$p \leftarrow 1$ \KwTo $q$}{
Candidate IBIs set, $I_{Cp} = I_{Sp} \cup I_{Mp} \cup I_{Op}$
\begin{align*}
\hat{j},\hat{k}, \hat{l} = \underset{ j \neq k \neq l}{\mathrm{argmin}} \{ |i_{Cp}^{j} - i_{Ap}^{1}|  &+|i_{Cp}^{k} - i_{Ap}^{2}| \\
 &+ |i_{Cp}^{l} - i_{Ap}^{3}|  \}
\end{align*}
\hspace{3pt} subject to $i_{Cp}^{j}, i_{Cp}^{k}, i_{Cp}^{l} \in I_{Cp}$\\
$I_{Fp} = \{$ $i_{Cp}^{\hat{i}}, i_{Cp}^{\hat{j}}, i_{Cp}^{\hat{k}}$ \} \\
$I_{F} = \hspace{2pt} I_{F} +\!\!\!+  I_{Fp}$
  }
    \textbf{return} $I_{F}$
 \end{algorithm2e}
}

\begin{table*}[]
\floatconts
  {tab:IBI_IEEE1}
  {\caption{IBI estimation using greedy-optimized approach on noisy PPGs from IEEE-Training dataset by adopting morphological features, systolic peaks (SP), maximum slope (MS), and onset.}}
  {  \vskip -0.2in
\resizebox{0.95\linewidth}{!}{
  \begin{tabular}{cccccc} 
    \toprule
    &  \multicolumn{2}{c}{One-channel(PPG1)} & One-channel(PPG2) & \multicolumn{2}{c}{Two-channel (PPG1\&2)} \\ 
\cmidrule(lr){2-3} \cmidrule(lr){4-4} \cmidrule(lr){5-6}
      & Aygun et al.(2019)   & My Result & My Result & Huang et al.(2021) &  My Result  \\
      &  Corr $|$ MAPE &  Corr $|$ MAPE & Corr $|$ MAPE & Corr $|$ MAPE  & Corr $|$ MAPE \\
    \midrule
    SP & 0.82 $|$ \hspace{0.1cm} n/a \hspace{0.2cm}  & 0.83 $|$ \hspace{0.1cm}8.5 \% & 0.78 $|$10.6 \%  & 0.90 $|$ \hspace{0.1cm}5.9 \% & 0.90 $|$ \hspace{0.1cm}5.9 \%\\
    MS & 0.85 $|$ \hspace{0.1cm} n/a \hspace{0.2cm}  & 0.83 $|$ \hspace{0.1cm}8.1 \% & 0.82 $|$ \hspace{0.1cm}9.3 \%  & 0.92 $|$ \hspace{0.1cm}4.8 \% & 0.92 $|$ \hspace{0.1cm}4.8 \%\\
    Onset & 0.86 $|$ \hspace{0.1cm} n/a \hspace{0.2cm}  & 0.86 $|$ \hspace{0.1cm}7.7 \% & 0.84 $|$ \hspace{0.1cm}8.5 \%  & 0.94 $|$ \hspace{0.1cm}4.5 \% & 0.94 $|$ \hspace{0.1cm}4.5 \%\\
    \midrule
   \hspace{0.4cm} \textbf{Fusion} \hspace{0.4cm} & 0.89 $|$ \hspace{0.1cm} n/a \hspace{0.2cm}  & $\textbf{0.96}^{*}$ $|$ $\textbf{3.2 \%}^{*}$ & $\textbf{0.95}^{*}$ $|$  $\textbf{3.7 \%}^{*}$  & n/a $|$ \hspace{0.1cm}n/a & $\textbf{0.98}^{*}$ $|$ $\textbf{2.2 \%}^{*}$\\
    \bottomrule
    \multicolumn{6}{l}{\small{
    $^{*}$ Results have significantly improvements compared to Huang et al. 2021 Onset ($p < 0.001$, paired t-test).}}\\
    \multicolumn{6}{l}{\small{
    Not available: n/a}}\\
    \end{tabular}
    }}
\end{table*}

\section{Dataset and Data Preprocessing}
The proposed techniques are evaluated on two datasets, the 2015 IEEE Signal Processing Cup training dataset (referred to as IEEE-Training) and the PPG-DaLiA dataset to have a comprehensive evaluation of performance during intensive exercises and daily activities~\citep{10,29}. The IEEE-Training dataset emphasizes the lab-based controlled conditions whereas the PPG-DaLiA dataset focuses on daily real-life activities.

\subsection{IEEE-Training Dataset.} Two-channel PPG signals (PPG1 and PPG2) from wrist-worn sensors and one-channel ECG signal were collected synchronously from 12 healthy individuals aged 18 to 35 while they were running on the treadmill with different speeds ~\citep{10}. The running program was set up as Rest 30s $\to$ Jogging 1 min $\to$ Running 1 min $\to$ Jogging 1 min $\to$ Running 1 min $\to$ Rest 30s. Both ECG and PPG signals are at a sampling rate of 125 Hz and upsampled to 500 Hz. Then, the single-channel PPG signals are preprocessed with a band-pass Butterworth filter with a cutoff frequency of 0.5 Hz and 15Hz whereas two-channel PPG signals are preprocessed with a band-pass Butterworth filter with a cutoff frequency of 0.7 Hz and 15Hz. ECG signals are filtered with a high-pass Butterworth filter with a 0.5 Hz cutoff frequency.

\subsection{PPG-DaLiA Dataset.} This dataset includes synchronized PPG and ECG signals recorded from wrist-worn devices (Empatica E4) and chest-worn devices (RespiBAN Professional), respectively~\citep{29}. Data was recorded from 15 subjects while performing different kinds of daily activities as naturally as possible for 2.5 hours. Two intense physical activities are selected, ascending/descending stairs (5 mins) and cycling (8 mins). The PPG signals from the PPG-DaLiA dataset are upsampled from 64 Hz to 500 Hz and filtered with a band-pass filter with a cutoff frequency of 0.5 Hz and 15Hz. The true R-peaks of ECG provided in this dataset are used to calculate the ground-truth IBIs for performance evaluation.

\begin{figure*}[h]
\vskip 0.0in
\begin{center}
\centerline{\includegraphics[width= 2 \columnwidth]{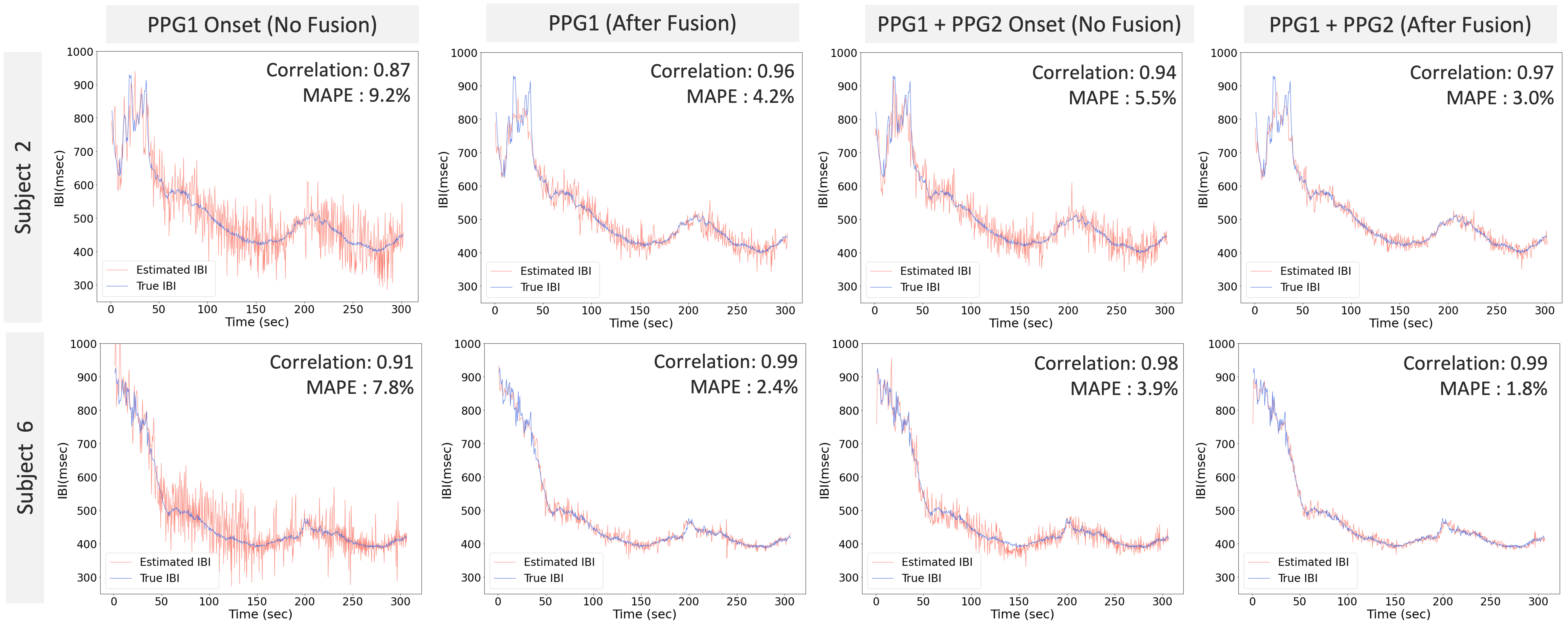}}
\end{center}
\vskip -0.4in
\caption{IBI sequence plots over time for subject 2 and subject 6 in IEEE-Training Dataset in comparison with or without the proposed greedy fusion method.}
\label{IBIplot}
\end{figure*}

\section{Results}
\subsection{Interbeat Intervals Evaluation}
This study evaluates the agreement between true and estimated IBIs using Pearson Correlation Coefficient (Corr) for each subject. As for accuracy performance metric, it relies on Mean Absolute Percentage Errors (MAPE) for each subject, defined as ~(\ref{MAPE}):
\begin{equation}
            \displaystyle
MAPE= \frac{1}{n} \sum_{i=1}^n \left[ \frac{|trueIBI_i- estimatedIBI_i |}{trueIBI_i} \times 100 \right] 
\label{MAPE}
\end{equation}
where $n$ is the total number of IBIs in a subject, $trueIBI_i$ denotes the $i$th true IBI from ECG signal and $estimatedIBI_i$ denotes the $i$th estimated IBI from PPG signal.

\subsubsection{Evaluation on IEEE-Training.}
Table~\ref{tab:IBI_IEEE1} depicts the overall performance evaluation of IBIs estimation on IEEE-Training dataset from single-channel and two-channel models using each morphological feature individually and fusion of them. The results are the average of 12 subjects in the IEEE-Training. As shown in Table~\ref{tab:IBI_IEEE1}, IBIs estimation in the one-channel PPG has significant improvements using the greedy fusion method as compared to the cases using each morphological feature individually. The correlation of one-channel PPG1 from onset feature without fusion is 0.86, whereas the correlation reaches to 0.96 after applying the greedy fusion method, which is improved by 11.6\%. Similarly for the one-channel PPG2, the correlation jumps greatly from 0.84 to 0.95, which is improved by 13.1\%. In addition, the MAPE is reduced to 3.2\% and 3.7\% in the one-channel PP1 and PPG2, respectively, which is 58.4\% and 56.5\% improvement as compared to the case using the onset feature individually.  If the greedy fusion method combines with the two-channel PPGs, it achieves correlation of 0.98 and MAPE of 2.2\%, where the MAPE is improved by 51.1\% (reducing from 4.5\% to 2.2\%) as compared to the case using the onset feature individually. In addition, the results of one-channel and two-channel PPGs using greedy fusion method have significant improvements compared to the best result of ~\citep{9}, which is onset from two-channel PPGs, with p-value $<$ 0.001 by paired t-test. The performance for individual subject are reported in Appendix~\ref{apd:second}.

The severely fluctuating patterns could be explicitly seen in estimated IBIs sequence calculated from the PPG1 onset feature of subject 2 and subject 6 in IEEE-Training, which is shown in Fig.~\ref{IBIplot}. The greedy fusion method mitigates overestimated fluctuating patterns of estimated IBIs sequence. In one-channel PPG1, it improves the correlation by 8.8\% and 10.3\% and reduces the MAPE by 69.2\% and 54.3\% in subject 6 and subject 2, respectively. Although two-channel PPGs provided high correlations of 0.98 and low percentage errors of 3.9\% in subject 6, the overestimated fluctuation challenge remains. The effectiveness of greedy fusion method is well demonstrated that the estimated IBIs are much closer to the true IBIs with low fluctuation. The correlations after applying greedy fusion method achieve 0.97 and 0.99 for subject 2 and subject 6, respectively, and MAPEs reduce to 3.0\% and 1.8\% in two-channel PPG signals.

\subsubsection{Evaluation on PPG-DaLiA.}
This approach is also applied to two daily activities, ascending/descending stairs (5 mins) and cycling (8 mins) on the PPG-DaLiA dataset. The PPG-DaLiA only provides one channel PPG from commercial wearable wristbands~\citep{29}. The average heart rate provided in this dataset is used to calculate average IBIs. The one-channel PPG with fusion of three morphological features achieves high correlation of 0.91 ± 0.04 and low MAPE of 3.8\% ± 0.8\% for ascending/descending stairs activity and high correlation of 0.95 ± 0.04 and low MAPE of 2.4\% ± 0.7\% for cycling activity, as shown in Table~\ref{tab:IBI_DaLiA}. The performance for individual subject are reported in Appendix~\ref{apd:second}

\begin{table}[h] 
    \centering
\caption{IBIs estimation performance of greedy-optimized approach using one-channel PPG in PPG-DaLiA dataset.}
\vskip -0.05in
\begin{tabular}{ccc} 
    \toprule
    & Stairs & Cycling \\
    & Corr $|$ MAPE &  Corr $|$ MAPE \\
    \midrule
    Average & 0.91 $|$ 3.8 \% & 0.95 $|$ 2.4 \% \\
    SD & 0.04 $|$ 0.8 \% & 0.04 $|$ 0.7 \% \\
    \bottomrule
    \end{tabular}
\label{tab:IBI_DaLiA}
\end{table}

\begin{table*}[h]
    \centering
\caption{HRV analysis based on results of IBI estimation using the proposed greedy-optimized approach in one-channel and two-channel PPG in IEEE-Training. All p-value of my Corr are less than 0.001. }
\vskip -0.05in
    \resizebox{0.9 \linewidth}{!}{
    \begin{tabular}{ cccc } 
    \toprule
           &  Aygun \textit{et al.} (PPG1) & My Result (PPG1) &  \hspace{0.3cm} My Result (PPG1\&2) \hspace{0.3cm} \\ 
        HRV Parameters & \hspace{0.45cm} Corr $|$ MAPE & \hspace{0.3cm}  Corr $|$ MAPE &  \hspace{0.3cm} Corr $|$ MAPE \\
    \midrule 
       Mean RR ($ms$)   &  0.986 $|$ n/a &  0.997 $|$ 0.6 \% &  0.999 $|$ 0.2 \% \\
       Mean HR ($1/min$) &  0.987 $|$ n/a &  0.995 $|$ 0.7 \% & 0.999 $|$ 0.2 \% \\
      SDNN ($ms$)  &  0.956 $|$ n/a &  0.996 $|$ 2.3 \% &   0.998 $|$ 1.3 \% \\
      STD HR ($1/min$)  &  0.860 $|$ n/a &  0.964 $|$ 4.5 \% &  0.990 $|$ 1.7 \% \\
      VLF Power ($ms^2$)   &   0.981 $|$ n/a &   0.993 $|$ 0.2 \% &   0.995 $|$ 0.2 \% \\
      LF Power ($ms^2$)  &   0.898 $|$ n/a &   0.971 $|$ 0.7 \% &   0.971 $|$ 0.8 \% \\
      HF Power ($ms^2$)   &   0.828 $|$ n/a &   0.851 $|$ 1.9 \% &   0.858 $|$ 1.7 \% \\
     \hspace{0.5cm} Total Power ($ms^2$) \hspace{0.5cm} &  0.974 $|$ n/a &  0.925 $|$ 1.1 \% &   0.932 $|$ 1.0 \% \\
    \bottomrule
    \end{tabular} 
    }
    \label{tab:HRV_IEEE}
\end{table*}

\subsection{Heart Rate Variability Analysis}
HRV parameters has been commonly used in the medical stratification of cardiac risk of morbidity and mortality for heart attack survivors ~\citep{HRVoverview}. The estimated and true IBIs are used to calculate time-domain and frequency-domain HRV parameters using pyHRV~\citep{32}. The time-domain HRV parameters investigated in this study include Mean RR, SDNN, Mean HR and STD HR. The frequency domain HRV parameters are computed using the autoregressive method to separate HRV into its component frequency band, including the VLF Power, LF Power, HF Power and the Total Power. The HRV analysis results are evaluated by Pearson Correlation Coefficient (Corr) and the accuracy are evaluated by Mean Absolute Percentage Errors (MAPE) for all subjects in the dataset, which is defined as ~(\ref{MAPE_HRV}).
 \begin{equation}
 \resizebox{.85\linewidth}{!}{$
             \displaystyle
 MAPE= \frac{1}{n} \sum_{i=1}^n \left[ \frac{|trueHRV_i- estHRV_i |}{trueHRV_i} \times 100 \right] $}
 \label{MAPE_HRV}
 \end{equation}
 where $n$ is the total number of subjects in the dataset, $trueHRV_i$  denotes the HRV parameter derived from true IBIs and $estHRV_i$ denotes the HRV parameter derived from estimated IBIs of the $i$’th subject.

Table~\ref{tab:HRV_IEEE} shows HRV analysis results on IEEE-Training dataset. My results are based on IBI estimation results of single-channel (PPG1) and two-channel (PPG1\&2) model after the greedy fusion method is applied on three morphological features. The result in \citet{19} was based on estimated IBIs from single-channel (PPG1) using their signal-switching fusion method of three morphological features. Results demonstrate the greedy fusion method has better performance in HRV parameters estimation. The estimated and true HRV parameters are highly correlated with low percentage errors both in one-channel and two-channel PPG signal. The Pearson correlation coefficients are above 0.93 significantly, except the coefficient of HF Power. All p-values of my Pearson Correlation Coefficient are less than 0.001. For two-channel PPG, MAPEs are all less than 1.7\% for eight HRV parameters. 

\begin{table}[h]
    \centering
\caption{HRV analysis based on results of IBIs estimation using the proposed greedy-optimized approach in single-channel PPG from Empatica E4 on PPG-DaLiA. All p-value of Corr are less than 0.001. 
} 
\vskip -0.05in
\resizebox{1 \linewidth}{!}{
    \begin{tabular}{ ccc } 
    \toprule
          & Stairs &  Cycling  \\ 
       HRV parameters & \hspace{0.3cm} Corr $|$ MAPE & \hspace{0.3cm} Corr $|$ MAPE\\
    \midrule 
      Mean RR ($ms$)  & \hspace{0.2cm} 1 \hspace{0.2cm} $|$ 0.2 \% &  \hspace{0.2cm} 1 \hspace{0.2cm} $|$ 0.1 \% \\
      Mean HR ($1/min$) &  \hspace{0.2cm} 1 \hspace{0.2cm} $|$ 0.2 \% &  \hspace{0.2cm} 1 \hspace{0.2cm} $|$ 0.1 \% \\
     SDNN ($ms$) & 0.981 $|$ 5.0 \% &  0.998 $|$ 2.4 \% \\
      STD HR ($1/min$) &0.935 $|$ 6.5 \% &  0.996 $|$ 2.7 \% \\
      VLF Power ($ms^2$)  &0.996 $|$ 0.2 \% &  0.998 $|$ 0.2 \% \\
      LF Power ($ms^2$) &  0.905 $|$ 1.1 \% &  0.965 $|$ 0.8 \% \\
      HF Power  ($ms^2$) &  0.786 $|$ 4.4 \% &  0.784 $|$ 3.5 \% \\
      Total Power ($ms^2$) &  0.843 $|$ 2.7 \% &  0.888 $|$ 1.9 \% \\
    \bottomrule
    \end{tabular} 
    } 
\label{tab:HRV_DaLiA}
\end{table}

HRV analysis of PPG-DaLiA Dataset is shown in Table~\ref{tab:HRV_DaLiA}, which is based on IBIs estimation results of one-channel PPG from Empatica E4 of 15 subjects in stairs and cycling activities after the greedy fusion method is applied on three features, systolic peaks, maximum slope and onset. Results show that the estimated and true HRV parameters are highly correlated with low absolute percentage errors for both stairs and cycling activities and the performance of the cycling activity is better than the stairs activity. Note that HF Power has the lowest correlation among the eight HRV parameters in the stairs and cycling activities of the PPG-DaLiA and in the IEEE-Training. HF Power has the highest absolute errors among the four frequency-domain parameters. 

\begin{figure}[h]
\vskip -0.0in
\begin{center}
\centerline{\includegraphics[width=1 \columnwidth]{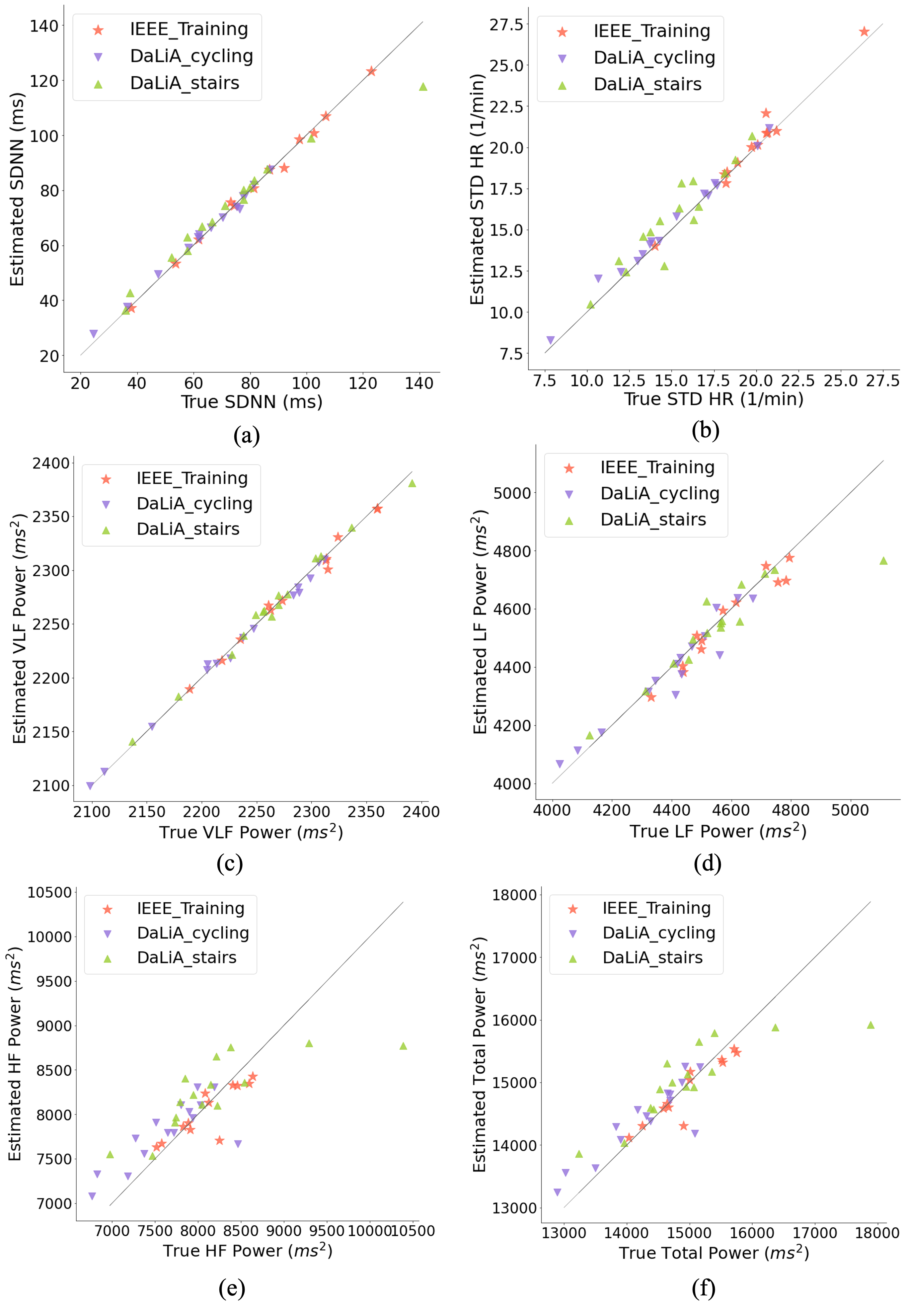}}
\end{center}
\vskip -0.5in
\caption{Scatterplot comparison of true/estimated HRV parameters (a) SDNN (b) STD HR (c) VLF Power (d) LF Power (e) HF Power and (f) Total Power from motion-contaminated PPG signals on IEEE-Training (running on treadmill) and PPG-DaLiA (stairs and cycling).}
\label{HRVCorrelation}
\end{figure}

Fig.~\ref{HRVCorrelation} provides the scatterplots that compare the true and estimated SDNN, STD HR, VLF Power, LF Power, HF Power and Total Power derived from motion-contaminated PPG signals during intensive treadmill activities on IEEE-Training dataset and in the stairs and cycling activities of PPG-DaLiA Dataset. These plotted points in Fig.~\ref{HRVCorrelation} (a, b, c, d, e, f) are distributed along with the identity line closely, showing that the true and estimated HRV results have high correlations and small absolute errors in all three intensive activities. HF Power, however, is often overestimated in the PPG-DaLiA.

\section{Discussion}
IBI/HRV estimation from PPG are challenging because motion-artifacts induced by intensive daily and exercise activities significantly deteriorates the accuracy. The common strategy in analyzing IBI and HRV from noisy PPG is discarding motion-contained signal segments, which loses potential health information triggered during exercise. IBI estimation from the two-channel PPG outperform the one-channel PPG in a previous study~\citep{9}. Nevertheless, the overestimated fluctuating patterns of estimated IBIs sequence have still remained and the method could not be adapted with prevailing single-contact PPG wearables. 

The proposed greedy-optimized approach could tackle this challenge with efficient time complexity of $O(n)$. Results from PPG-DaLiA indicate that my method can accurately estimate IBIs and HRV from one-channel PPG, which can be embedded with single-contact commercial wearables. Average HR, however, is an important input. The accuracy of IBI/HRV estimation would be limited when the accuracy of average HR decreases. Favorably, this greedy-optimized approach for IBI/HRV estimation is independent of average HR. Any algorithm that generates accurate average HR from noise-contaminated PPG could be applied, such as WFPV~\citep{11} and Deep PPG~\citep{29}. Another attention that this method is evaluated on the PPG with 5-8 minutes duration. Future work can extend to long-term period wearable PPG signals.

\section{Conclusion }
In this study, a greedy-optimized approach provides high accuracy in estimating IBI/HRV from PPG signals obtained during intensive daily activities. It exploits a convex penalty function to optimize weight assignment in the shortest path calculation and a greedy fusion method to select optimal IBIs in each step. In the 2015 IEEE Signal Processing Cup, the approach achieves low average percentage errors of 2.2\% and 3.2\% with high correlations of 0.98 and 0.96 for IBIs estimation through two-channel PPGs and one-channel PPG1, respectively. The estimated and true HRV parameters are highly correlated with low percentage error. This greedy-optimized approach is further validated in daily activities on the PPG-DaLiA. The estimated IBIs achieve high correlations of 0.91 and 0.95 with low percentage error of 3.8\% and 2.4\% for stairs and cycling activities, indicating the robustness of proposed techniques.

\clearpage

\appendix
\section{IBI Estimation for Each Subject in IEEE\_Training dataset and PPG-DaLiA dataset}\label{apd:second}

\begin{table}[h]
\centering
\caption{IBI estimation performance using greedy-optimized approach of PPG signals on IEEE\_Training dataset.}
\vskip -0.05in
    \resizebox{1\linewidth}{!}{
    \begin{tabular}{ccccc} 
    \toprule
    Subject  & PPG1 & PPG2 & PPG1\&2  \\
      ID &  Corr $|$ MAPE & Corr $|$ MAPE & Corr $|$ MAPE   \\
    \midrule
    1 & 0.98 $|$ \hspace{0.1cm} 3.4 \% \hspace{0.2cm} & 0.98 $|$ \hspace{0.1cm}4.4 \% & 0.99 $|$ \hspace{0.1cm}2.5 \%  \\
    2 & 0.96 $|$ \hspace{0.1cm} 4.2 \% \hspace{0.2cm} & 0.94 $|$ \hspace{0.1cm}4.9 \% & 0.97 $|$ \hspace{0.1cm}3.0 \%  \\
    3 & 0.96 $|$ \hspace{0.1cm} 3.5 \% \hspace{0.2cm} & 0.95 $|$ \hspace{0.1cm}3.9 \% & 0.99 $|$ \hspace{0.1cm}2.0 \%  \\
    4 & 0.97 $|$ \hspace{0.1cm} 3.1 \% \hspace{0.2cm} & 0.96 $|$ \hspace{0.1cm}3.7 \% & 0.98 $|$ \hspace{0.1cm}2.0 \%  \\
    5 & 0.98 $|$ \hspace{0.1cm} 2.0 \% \hspace{0.2cm} & 0.97 $|$ \hspace{0.1cm}2.5 \% & 0.99 $|$ \hspace{0.1cm}1.5 \%  \\
    6 & 0.99 $|$ \hspace{0.1cm} 2.4 \% \hspace{0.2cm} & 0.99 $|$ \hspace{0.1cm}2.8 \% & 0.99 $|$ \hspace{0.1cm}1.8 \%  \\
    7 & 0.98 $|$ \hspace{0.1cm} 2.0 \% \hspace{0.2cm} & 0.97 $|$ \hspace{0.1cm}2.9 \% & 0.99 $|$ \hspace{0.1cm}1.6 \%  \\
    8 & 0.98 $|$ \hspace{0.1cm} 2.7 \% \hspace{0.2cm} & 0.96 $|$ \hspace{0.1cm}4.0 \% & 0.99 $|$ \hspace{0.1cm}1.9 \%  \\
    9 & 0.99 $|$ \hspace{0.1cm} 2.4 \% \hspace{0.2cm} & 0.98 $|$ \hspace{0.1cm}3.1 \% & 0.99 $|$ \hspace{0.1cm}1.7 \%  \\
    10 & 0.83 $|$ \hspace{0.1cm} 4.6 \% \hspace{0.2cm} & 0.86 $|$ \hspace{0.1cm}4.0 \% & 0.93 $|$ \hspace{0.1cm}2.9 \%  \\
    11 & 0.91 $|$ \hspace{0.1cm} 3.2 \% \hspace{0.2cm} & 0.86 $|$ \hspace{0.1cm}5.1 \% & 0.93 $|$ \hspace{0.1cm}2.8 \%  \\
    12 & 0.95 $|$ \hspace{0.1cm} 4.5 \% \hspace{0.2cm} & 0.97 $|$ \hspace{0.1cm}3.2 \% & 0.98 $|$ \hspace{0.1cm}2.5 \%  \\
    \midrule
    Average & 0.96 $|$ \hspace{0.1cm} 3.2 \% \hspace{0.2cm} & 0.95 $|$ 3.7 \% & 0.98 $|$  2.2 \%  \\
    SD & 0.04 $|$ \hspace{0.1cm} 0.9 \% \hspace{0.2cm} & 0.04 $|$ 0.8 \% & 0.02 $|$  0.5 \%  \\
    \bottomrule
    \end{tabular}
    }
\label{tab:IBI_IEEE_breakdown}
\end{table}

\begin{table}[h]
\centering
\caption{IBI estimation performance of greedy-optimized approach using one-channel PPG in the PPG-DaLiA dataset.}
\vskip -0.05in
    \resizebox{0.9\linewidth}{!}{
    \begin{tabular}{ccc} 
    \toprule
    Subject  & Ascending/Descending & Cycling   \\
     ID & Stairs (5mins) & (8mins)   \\
      &  Corr $|$ MAPE & Corr $|$ MAPE   \\
    \midrule
    1 & 0.96 $|$ \hspace{0.1cm} 3.7 \% \hspace{0.2cm} & 0.97 $|$ \hspace{0.1cm}1.9 \%  \\
    2 & 0.94 $|$ \hspace{0.1cm} 3.9 \% \hspace{0.2cm} & 0.94 $|$ \hspace{0.1cm}3.1 \%  \\
    3 & 0.92 $|$ \hspace{0.1cm} 2.8 \% \hspace{0.2cm} & 0.97 $|$ \hspace{0.1cm}2.3 \%   \\
    4 & 0.91 $|$ \hspace{0.1cm} 4.4 \% \hspace{0.2cm} & 0.94 $|$ \hspace{0.1cm}3.5 \%   \\
    5 & 0.85 $|$ \hspace{0.1cm} 4.3 \% \hspace{0.2cm} & 0.87 $|$ \hspace{0.1cm}2.6 \%   \\
    6 & 0.89 $|$ \hspace{0.1cm} 2.8 \% \hspace{0.2cm} & 0.97 $|$ \hspace{0.1cm}1.6 \% \\
    7 & 0.89 $|$ \hspace{0.1cm} 4.2 \% \hspace{0.2cm} & 0.97 $|$ \hspace{0.1cm}2.2 \%   \\
    8 & 0.85 $|$ \hspace{0.1cm} 5.3 \% \hspace{0.2cm} & 0.89 $|$ \hspace{0.1cm}2.7 \%   \\
    9 & 0.96 $|$ \hspace{0.1cm} 3.8 \% \hspace{0.2cm} & 0.98 $|$ \hspace{0.1cm}2.9 \%   \\
    10 & 0.87 $|$ \hspace{0.1cm} 4.6 \% \hspace{0.2cm} & 0.92 $|$ \hspace{0.1cm}3.2 \%  \\
    11 & 0.91 $|$ \hspace{0.1cm} 3.8 \% \hspace{0.2cm} & 0.99 $|$ \hspace{0.1cm}1.5 \%  \\
    12 & 0.92 $|$ \hspace{0.1cm} 4.0 \% \hspace{0.2cm} & 0.98 $|$ \hspace{0.1cm}1.8 \%  \\
    13 & 0.98 $|$ \hspace{0.1cm} 2.4 \% \hspace{0.2cm} & 0.99 $|$ \hspace{0.1cm}1.5 \%  \\
    14 & 0.96 $|$ \hspace{0.1cm} 2.6 \% \hspace{0.2cm} & 0.98 $|$ \hspace{0.1cm}2.2 \%  \\
    15 & 0.90 $|$ \hspace{0.1cm} 4.2 \% \hspace{0.2cm} & 0.91 $|$ \hspace{0.1cm}3.5 \%  \\
    \midrule
    Average & 0.91 $|$ \hspace{0.1cm} 3.8 \% \hspace{0.2cm} & 0.95 $|$ 2.4 \%   \\
    SD & 0.04 $|$ \hspace{0.1cm} 0.8 \% \hspace{0.2cm} & 0.04 $|$ 0.7 \% \\
    \bottomrule
    \end{tabular}
    } 
\label{tab:IBI_DaLiA_breakdown}
\end{table}

\newpage

Table~\ref{tab:IBI_IEEE_breakdown} presents the IBI estimation for each subject on IEEE\_Training dataset using the greedy-optimized approach on PPG sigals. Table~\ref{tab:IBI_DaLiA_breakdown} presents the IBI estimation for each subject on PPG-DaLiA dataset using the greedy-optimized approach. Note that the true IBIs annotation of subject 10 stairs activity and subject 6 cycling activity are corrected by filtering out the abnormal high spikes. The estimated IBIs using my greedy-optimized approach are highly correlated with the true IBIs for each subject with MAPE less than 5 \% for all subjects on these two datasets.

\end{document}